# Hyperbolic Metamaterials for Enhanced Scintillation

**Priyankar Pandey[1], Subrahmanyam S G Mantha[1,2], Harish N S Krishnamoorthy[1*]**
[1]*Tata Institute of Fundamental Research Hyderabad, Hyderabad, Telangana 500046, India*
[2] *Birla Institute of Technology and Science Pilani, KK Birla Goa Campus, Zuarinagar, Goa 403726, India*
**Correspondence: harishk@tifrh.res.in*

**Abstract:**

***Scintillators are indispensable for the detection of X-rays, γ-rays, and energetic particles in applications ranging from medical imaging and security screening to high-energy physics. A major limitation of conventional scintillators is the inefficient extraction of scintillation photons caused by isotropic emission and total internal reflection at the scintillator–detector interface. Here, we propose and numerically investigate a grating-coupled hyperbolic metamaterial (HMM) in which the scintillation medium $CsPbBr_3$ forms one of the constituent layers. The proposed architecture simultaneously enhances spontaneous emission through broadband Purcell enhancement enabled by high-k modes and improves optical extraction by directing the emitted radiation along the hyperbolic resonance cone, with a bullseye grating providing efficient far-field outcoupling. Finite-difference time-domain simulations predict broadband enhancement of the detected scintillation signal exceeding threefold across the $CsPbBr_3$ emission band, with a peak enhancement of approximately fivefold around the emission wavelength of 523 nm compared with an equivalent bulk scintillator. The proposed multilayer architecture is compatible with established thin-film deposition and nanofabrication techniques, robust against fabrication-induced layer-thickness variations, and scalable to accommodate larger volumes of scintillating material, providing a practical route toward high-performance scintillators for radiation detection and imaging applications.***

## 1. Introduction

Scintillation is one of the most common techniques employed for the detection of energetic particles. It is a process through which certain materials emit visible light when irradiated with high-energy radiation such as X-rays, γ-rays or charged particles.[1] Scintillators are crucial to a wide range of technologies encompassing medical imaging, radiation therapy, non-destructive testing, security screening, environmental radiation monitoring, space technologies, and high energy particle detection. Conversion of high energy radiation into visible light in a scintillator occurs through three major steps – (i) interaction of the incident radiation with the scintillator deposits energy through processes such as the photoelectric effect or Compton scattering, producing energetic secondary electrons; (ii) these secondary electrons in turn create electron-hole pairs that propagate to luminescence centres within the material; (iii) each pair recombines radiatively, releasing energy as visible photons via spontaneous emission. The detected scintillation light yield is governed by two major factors: (i) the intrinsic photon generation efficiency of the scintillator and (ii) the efficiency with which the emitted photons are extracted and collected by the photodetector. Typical values of the latter in commercial scintillators are less than 20%, owing to loss factors such as total internal reflection at the scintillator-detector interface and isotropic emission of light.[2] Intrinsic scintillation efficiency has primarily been improved through the development of new scintillator materials, with photonic engineering emerging as a complementary strategy, whereas enhancement of light extraction efficiency has largely relied on engineering the photonic environment.[3,4] Nanostructuring scintillation materials into two-dimensional photonic crystals has been shown to enhance the light extraction efficiency.[5] Recently, it was proposed and experimentally demonstrated that engineering the band dispersion of a 1D photonic crystal enables improvement in both the extraction efficiency by directing the scintillated emission as well as enhancing the scintillation rate through Purcell effect.[6,7] Ye et al. demonstrated enhancement in light yield due to Purcell effect by using nanoscale-confined surface plasmon polariton modes at the interface of a metal-scintillator thin film system.[8] Despite these advances, achieving strong enhancement of both photon generation and optical extraction within a single photonic platform remains an outstanding challenge. In this work, we show that a grating-coupled hyperbolic metamaterial (HMM) incorporating the scintillation medium as one of its constituent layers substantially enhances the detected photon yield through two synergistic mechanisms: improved optical outcoupling via the hyperbolic resonance cone and broadband Purcell enhancement of spontaneous emission, resulting in up to a fivefold enhancement of the detected scintillation signal. We further show that the enhancement is robust against fabrication-induced layer-thickness variations and that the proposed architecture can be scaled to accommodate a larger volume of scintillating material.

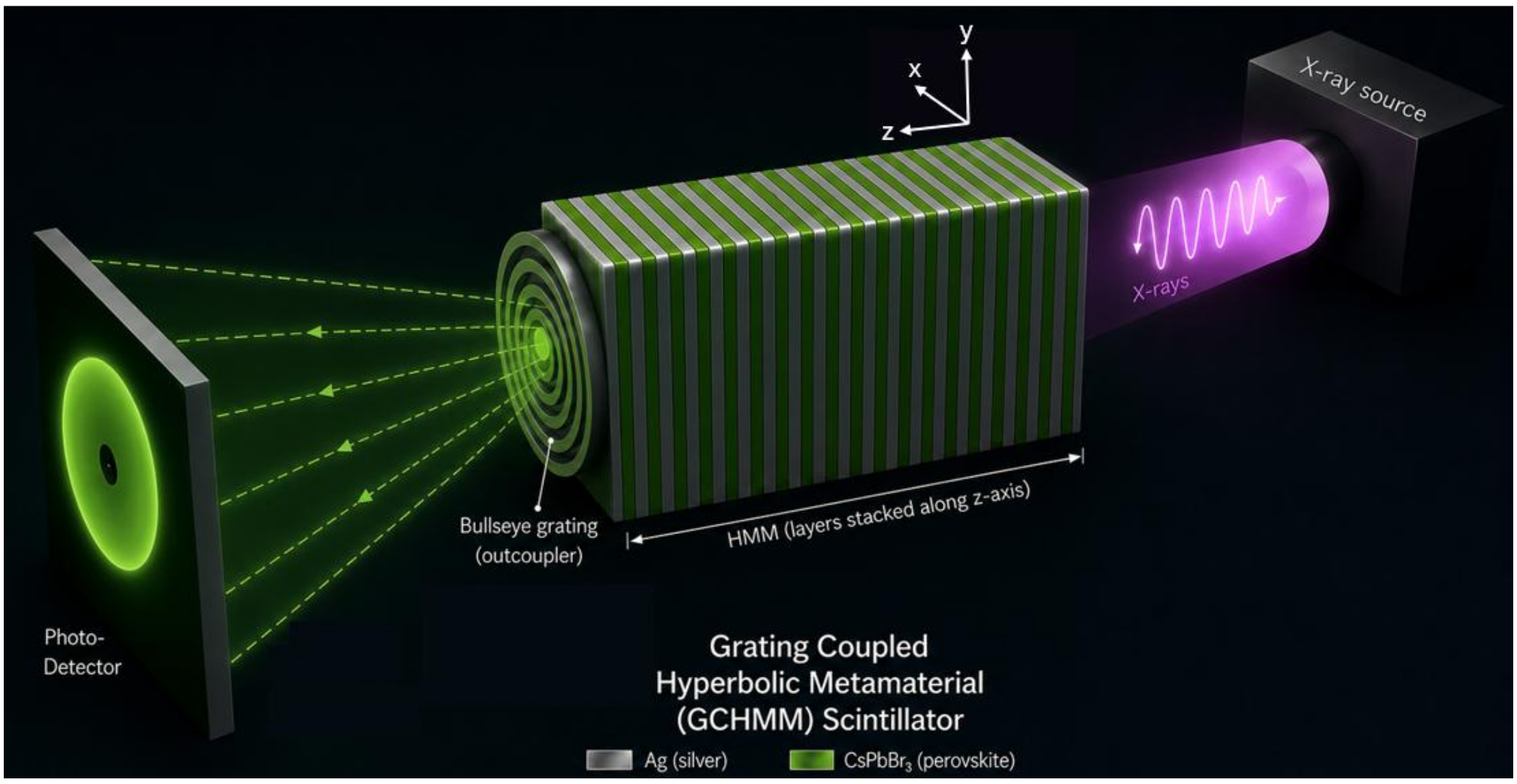

***Figure 1.*** *Conceptual illustration of the proposed grating-coupled hyperbolic metamaterial (GCHMM) scintillator. The HMM consists of alternating Ag and $CsPbBr_3$ layers, with the scintillation medium incorporated directly into the metamaterial. X-ray excitation generates visible emission within the $CsPbBr_3$ layers, which couples to the high-k states of the HMM. The resulting enhancement of the local density of optical states increases the spontaneous-emission rate (Purcell enhancement), while the hyperbolic resonance cone redistributes the emitted radiation into a directional wavefront. A germanium bullseye grating subsequently outcouples*

*these otherwise non-radiative high-k modes into free space, leading to enhanced photon collection by the photodetector.*

## 2. Results and Discussion

HMMs composed of subwavelength thick metal-dielectric layers are uniquely characterized by their ability to support high-*k* states (large in-plane wavevector modes), arising from coupling of surface plasmon polariton modes at successive metal–dielectric interfaces.[9–11] Due to the presence of the high-k states as well as the resulting enhancement in the photon density of states, they have been shown to be useful for a variety of applications such as imaging beyond the diffraction limit,[12–14] enhancing spontaneous emission rates of quantum emitters including perovskite nanocrystals,[15–23] developing highly sensitive sensing platforms,[24] nanoscale optical cavities,[25] and thermal energy harvesting.[26] Figure 1 shows a concept schematic of the proposed metamaterial-enhanced scintillator which uses a hyperbolic metamaterial (HMM) composed of alternating layers of silver (Ag) and scintillating metal-halide perovskite $CsPbBr_3$. A bullseye grating placed on top of the HMM structure facilitates the outcoupling of the otherwise non-radiative high-*k* modes into propagating far-field radiation. Upon X-ray irradiation, visible scintillation photons are generated within each perovskite layer. The presence of the high-*k* modes in the HMM substantially increases the local photonic density of states experienced by dipoles in the perovskite layers, thereby enhancing their spontaneous emission rate through the Purcell effect. In addition, the hyperbolic dispersion naturally channels electromagnetic energy along well-defined resonance cones, providing an efficient mechanism for directional extraction of the emitted photons.

**Figure 2a** shows a schematic representation of the proposed HMM design composed of 7 periods of alternate Ag and $CsPbBr_3$ layers of thicknesses 16 nm and 20 nm, respectively. $CsPbBr_3$ is chosen as the scintillation medium owing to its fast response time, high light yield, and excellent scintillation efficiency compared to conventional scintillators,[27–33] properties that are crucial for high-sensitivity radiation detection by enhancing signal strength and enabling detection of low-dose radiation. In addition, its emission spectrum centred around 523 nm, aligns well with the sensitivity range of widely used photodetectors such as silicon photomultipliers (SiPMs) and photomultiplier tubes (PMTs). This compatibility enhances detection efficiency and simplifies system integration.[30] The germanium bullseye grating aids in outcoupling the scintillation emission coupled to the high-k states of the HMM into the far-field which will otherwise stay confined within the HMM due to momentum mismatch between

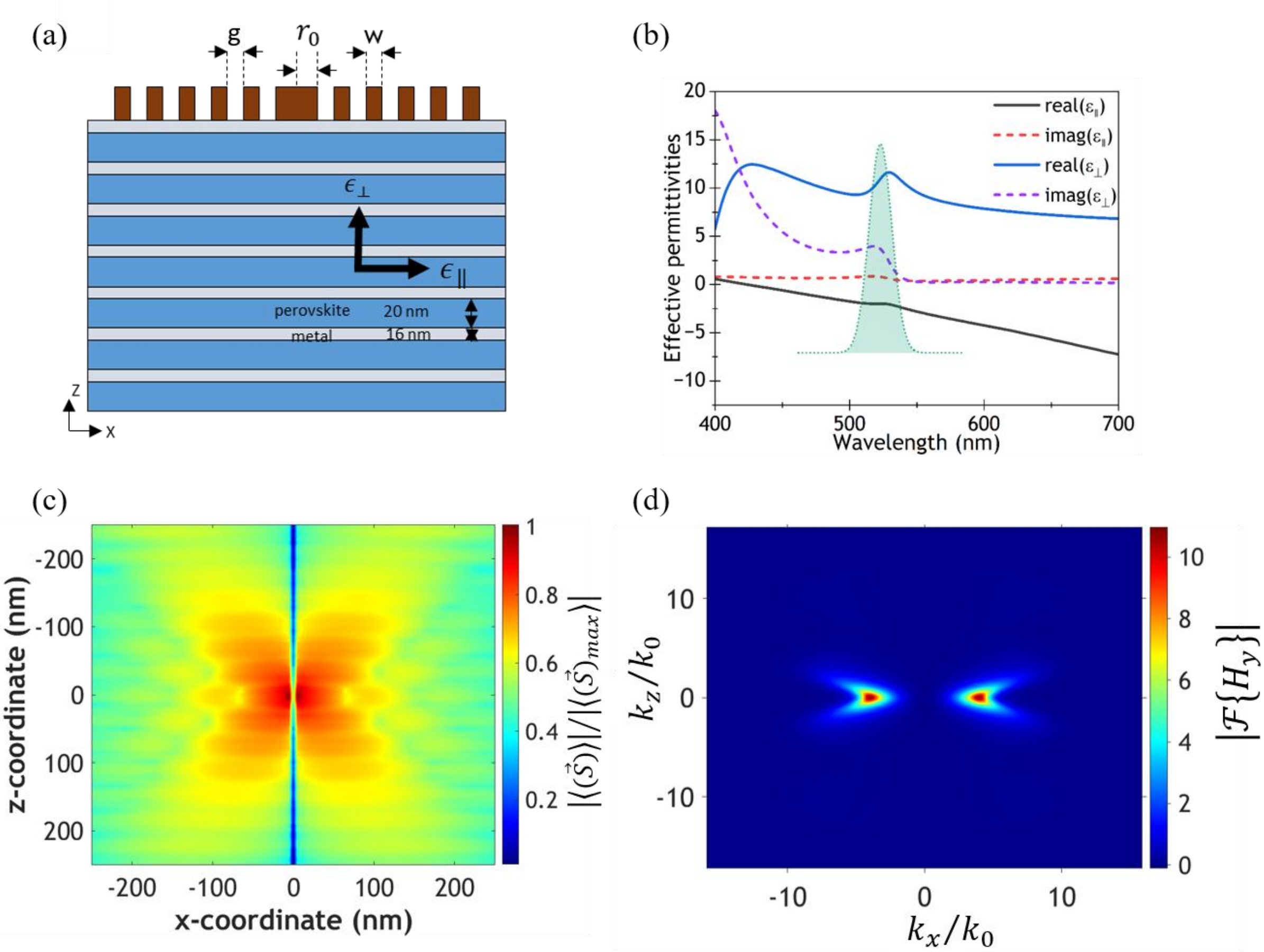

***Figure 2.*** *Design and electromagnetic characteristics of the proposed hyperbolic metamaterial (HMM) scintillator. (a) Schematic of the HMM consisting of seven Ag (16 nm)/$CsPbBr_3$ (20 nm) bilayers capped with a germanium bullseye grating for outcoupling the otherwise non-radiative high-k modes into free space. The effective dielectric tensor components parallel ($\varepsilon_\parallel$) and perpendicular ($\varepsilon_\perp$) to the layers are indicated. (b) Effective permittivity components of the HMM calculated using effective-medium theory. The multilayer is designed such that the $CsPbBr_3$ emission spectrum (green shaded region) lies within the Type-II hyperbolic regime,*

*where $Re(\varepsilon_{\parallel}) < 0$ and $Re(\varepsilon_{\perp}) > 0$. The $CsPbBr_3$ emission spectrum is shown on an arbitrary normalized scale for comparison. (c) Real-space distribution of the normalized magnitude of the time-averaged Poynting vector generated by a vertically oriented dipole embedded within the HMM, showing the characteristic X-shaped hyperbolic resonance cone and directional energy transport. (d) Magnitude of the two-dimensional spatial Fourier transform of the magnetic field, $|\mathcal{F}\{H_y(x,z)\}|$ at $\lambda = 523nm$, corresponding to the optical isofrequency contour of the multilayer. The hyperbolic nature of the isofrequency contour confirms coupling of the dipole emission to the high-k states supported by the HMM.*

the modes in the HMM and air. The dimensions of the grating, gap g, radius $r_0$, width w and thickness t are 120 nm, 43 nm, 144 nm and 58 nm, respectively which were obtained by maximizing the optical power (magnitude of surface integral of the Poynting vector) at the central emission wavelength of $CsPbBr_3$ reaching the detector positioned 1 µm above structure.

Since the layer thicknesses are deeply subwavelength, the HMM's optical properties can be quantified by employing effective medium approximation in terms of two effective permittivities, parallel ($\varepsilon_{\parallel}$) and perpendicular ($\varepsilon_{\perp}$) to the plane of the layers. **Figure 2b** shows the dispersion of the real and imaginary parts of $\varepsilon_{\parallel}$ and $\varepsilon_{\perp}$ retrieved by employing effective medium approximation. Permittivities of Ag and $CsPbBr_3$ retrieved from literature used for this calculation are plotted in the Supporting Information (Figure S1).[34,35] The thickness of the silver and perovskite layers are chosen such that the photoluminescence (PL) emission of the perovskite is in the Type II hyperbolic dispersion regime [$real(\varepsilon_{\parallel}) < 0;\ real(\varepsilon_{\perp}) > 0$ ] of the HMM which onsets from λ ≈ 420 nm. This ensures that the perovskite emission lies in the hyperbolic dispersion regime and enables it to be coupled to the high-k states of the HMM. While effective medium approximation provides an indicative idea of whether the perovskite emission couples to the high-k states, we confirm this by visualizing the electromagnetic fields of the layered structure in the real and Fourier space. We placed a vertically oriented electric dipole, with emission modelled as a Gaussian spectrum (523 nm centre, 20 nm FWHM, representative of $CsPbBr_3$ PL)[36] at the centre of the fifth perovskite layer, and computed the spatial distribution of the magnitude of the time-averaged Poynting vector within the metamaterial (**Figure 2c**). This distribution of the optical power density exhibits the characteristic X-shaped (bowtie) resonance-cone pattern that is a hallmark of hyperbolic metamaterials. To further verify the hyperbolic nature of the supported modes, we computed the optical isofrequency surface by taking the two-dimensional Fourier transform of the y-component of the magnetic field, $H_y$(**Figure 2d**). The resulting $k$--space distribution exhibits the characteristic hyperbolic isofrequency contour, in excellent agreement with the predictions of the effective-medium model. These results demonstrate that the dipole emission is (i) highly directional because of the hyperbolic dispersion and (ii) efficiently coupled to the high-$k$ states supported by the multilayer, providing the physical basis for the observed Purcell enhancement.[11,16,37] We would like to note that in this work the effective-medium model is used only to identify the spectral regime supporting hyperbolic dispersion; all electromagnetic simulations for quantifying scintillator performance were carried out using the actual multilayer geometry. To further verify that the HMM supports strong spontaneous-emission enhancement, the Purcell factor for a single vertically oriented dipole embedded within the multilayer was calculated (Figure S2, Supporting Information). The large Purcell factors obtained across the $CsPbBr_3$ emission band confirm the substantial enhancement of the local density of optical states supported by the HMM. We note that this quantity characterizes the local spontaneous-emission enhancement of an individual emitter and is therefore not directly comparable to the overall scintillation enhancement discussed later, which additionally incorporates optical absorption, finite grating outcoupling efficiency, and ensemble averaging over multiple emitters.

To quantify the combined influence of Purcell enhancement and improved optical outcoupling on the detected scintillation signal, we compute the optical power density reaching a detector placed above the metamaterial. The detector is represented by a surface monitor positioned 1 μm above the HMM, approximating the detector geometry used in conventional scintillation systems. The scintillation emission from each $CsPbBr_3$ layer is modelled as an ensemble of incoherent electric dipoles with random spatial positions and phases, representing the stochastic nature of radiative recombination following X-ray excitation (see Methods for details). All dipoles are considered to be vertically oriented so that their predominantly TM-polarized emission efficiently excites the coupled surface plasmon polariton modes supported by the HMM (see Methods). Such preferential dipole orientation can, in principle, be achieved through appropriate passivation of the perovskite layers.[38]

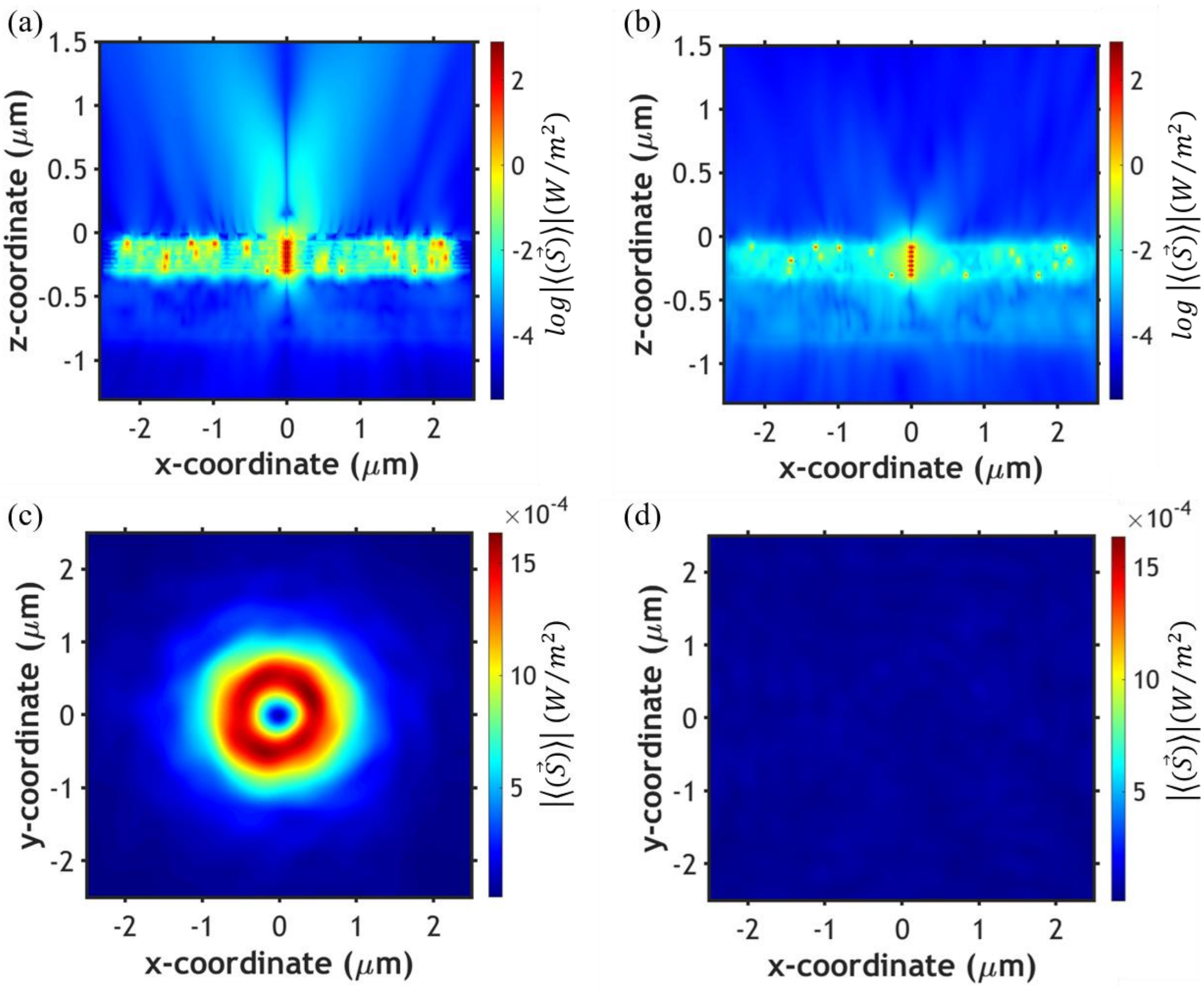

***Figure 3.*** *Colour maps of the magnitude of the time-averaged Poynting vector at the $CsPbBr_3$ emission wavelength (λ = 523 nm). (a,b) Spatial distribution in the xz plane for the (a) HMM scintillator and (b) reference bulk $CsPbBr_3$ scintillator under identical dipole excitation conditions. The HMM exhibits strongly directional emission arising from the hyperbolic resonance cone, whereas the bulk scintillator shows weak, diffuse radiation. (c,d) Optical power density (magnitude of the time-averaged Poynting vector) at the detector plane located at z = 1 μm above the structure for the (c) HMM and (d) bulk configurations. The HMM produces a pronounced ring-shaped intensity distribution, characteristic of*

*conical emission from the hyperbolic resonance cone, and delivers substantially greater optical power to the detector than the bulk scintillator.*

**Figures 3a** and **3b** plot the optical power density in the plane transverse to the detector for the case of the HMM scintillator and the bulk scintillator which is a perovskite film with thickness equal to the cumulative thickness of the HMM with identical dipole configuration. The HMM configuration delivers substantially higher optical power to the detector than the bulk scintillator. The enhanced directionality of the emitted radiation is also apparent in the HMM configuration. This is substantiated further by **Figures 3c** and **3d** which plot optical power density at the plane of the detector and reveal that the HMM-enhanced configuration yields a significantly larger intensity reaching the detector, whereas the bulk configuration yields minimal intensity. Notably, the HMM configuration produces a ring-shaped intensity pattern at the detector plane, consistent with the conical emission geometry imposed by the hyperbolic resonance cone, in contrast to the featureless, extremely low-intensity signal for the bulk case. Similar directional emission was observed for simulations involving a single dipole (Figure S3, Supporting Information). Incorporating multiple dipoles with randomized positions and phases slightly broadens the ring-shaped intensity profile because of the incoherent superposition of the individual emitters while preserving the overall directional emission characteristics.

While Figure 3 qualitatively demonstrates the enhanced light extraction provided by the grating-coupled HMM, a quantitative metric is required to evaluate the overall improvement in scintillation performance. The enhancement arises from the combined action of two intrinsic properties of the HMM: (i) enhancement of the spontaneous emission rate through the Purcell effect, enabled by the large local density of optical states associated with the high-$k$ modes, and (ii) directional redistribution of the emitted radiation through the hyperbolic resonance cone. The bullseye grating subsequently converts these otherwise non-radiative high-$k$ modes into propagating far-field radiation. Since these mechanisms operate simultaneously, their combined effect is quantified through the far-field optical power reaching the detector. Accordingly, the time-averaged Poynting vector integrated over the detector plane, $P_{HMM}$ serves as a direct measure of the detected scintillation power. We define the scintillation enhancement factor

$$\eta = \frac{P_{HMM}}{P_{Bulk}}$$

where the bulk reference $P_{Bulk}$ corresponds to power detected for a homogeneous $CsPbBr_3$ film having the same overall thickness and identical dipole configuration. This enhancement factor represents the overall improvement in the detected scintillation signal and therefore incorporates spontaneous emission enhancement, optical outcoupling, transmission through the multilayer–grating structure, and absorption losses, rather than the Purcell enhancement alone. **Figure 4a** plots the scintillation enhancement factor, η, as a function of wavelength at the detector location (z = 1 μm) and reveals a peak enhancement of approximately fivefold near 525 nm, close to the $CsPbBr_3$ emission peak at 523 nm and the grating design wavelength. Away from this spectral region, the enhancement decreases because the bullseye grating is optimized for efficient momentum matching at 523 nm, resulting in progressively lower outcoupling efficiency at wavelengths further from this design point. The small (~2 nm) offset between the enhancement maximum and the nominal design wavelength arises because η is determined by the combined wavelength dependence of the HMM emission, grating outcoupling, and the bulk reference signal ($P_{Bulk}$; Figure S5), rather than by the grating response alone. The apparent asymmetry in the enhancement spectrum arises primarily from the wavelength dependence of the bulk reference signal, which becomes very small at shorter wavelengths. Consequently, the ratio $\eta$ is artificially amplified in this spectral region. This interpretation is supported by the individual spectra of $P_{Bulk}$ and $P_{HMM}$ (Supporting information Figure S5) and by the corresponding single-dipole calculations (Supporting information Figure S4), which exhibit a more symmetric spectral response. As expected, the calculated scintillation enhancement is significantly smaller than the Purcell factor as the former reflects the combined influence of spontaneous emission enhancement, absorption losses, finite outcoupling efficiency, and ensemble averaging over randomly distributed emitters.

In addition to enhancing the detected power, the HMM substantially modifies the angular distribution of the emitted radiation. **Figure 4b** shows that the emission is concentrated within a narrow angular range centered on the resonance-cone direction, whereas the bulk scintillator exhibits nearly isotropic emission. This pronounced directionality originates from the hyperbolic isofrequency contour of the HMM, which aligns the energy flow (Poynting vector) along the resonance cone before the bullseye grating outcouples it into free space. Together, the spectral enhancement and directional emission confirm that the grating-coupled HMM simultaneously increases spontaneous emission and improves optical extraction, leading to a substantial enhancement in the detected scintillation signal compared with a conventional bulk scintillator.

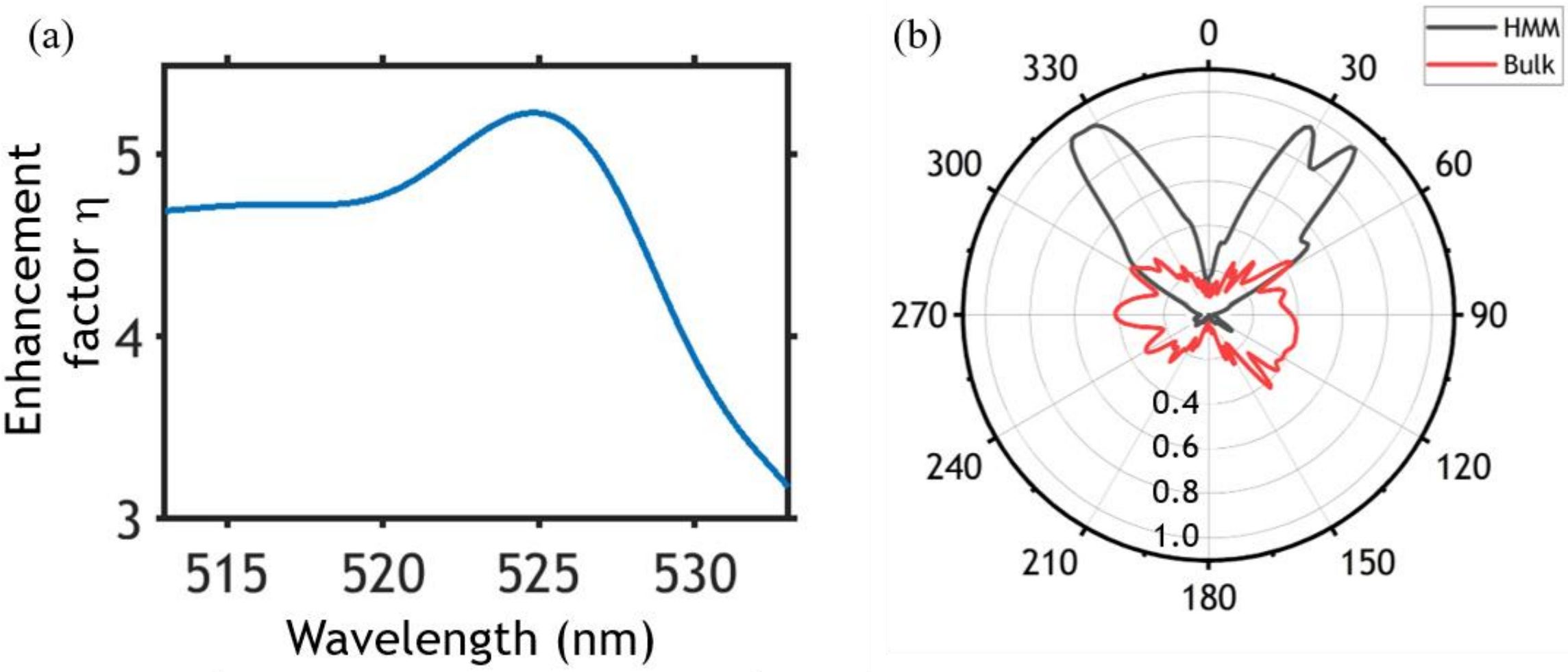

***Figure 4.*** *(a) Spectral variation of the scintillation enhancement factor,* $\eta = P_{HMM}/P_{Bulk}$*, evaluated at the detector plane located 1 μm above the structure. The enhancement reaches approximately fivefold near the* $CsPbBr_3$ *emission peak (523 nm), demonstrating the combined*

*effects of Purcell-enhanced spontaneous emission, directional energy transport through the hyperbolic resonance cone, and efficient grating-assisted outcoupling. (b) Normalized angular distribution of the emitted optical intensity for the grating-coupled HMM and the reference bulk scintillator. The HMM exhibits pronounced directional emission, whereas the bulk scintillator shows a comparatively weak and nearly isotropic radiation pattern. The two dominant emission lobes correspond to the intersection of the hyperbolic resonance cone with the observation plane and are a direct consequence of the anisotropic energy flow in the HMM.*

Previous studies have demonstrated scintillation enhancement using photonic platforms such as dispersion-engineered photonic crystals and plasmonic structures that rely on resonant coupling to surface plasmon polaritons. For example, Ye *et al.* demonstrated a 2.5-fold scintillation enhancement in a plasmonic structure comprising a 50 nm-thick $BA_2PbBr_4$ film, where Purcell enhancement mediated by surface plasmon polaritons was identified as the primary mechanism responsible for the enhanced scintillation yield.[8] In contrast, the proposed hyperbolic metamaterial (HMM) supports a continuum of coupled high-*k* electromagnetic states distributed throughout the multilayer rather than localized resonances confined to a single metal–dielectric interface. Consequently, spontaneous-emission enhancement is intrinsically broadband instead of being limited to a narrow spectral resonance. This behaviour is reflected in the calculated scintillation enhancement, which exceeds a factor of three across the entire $CsPbBr_3$ emission bandwidth and reaches approximately five at the emission peak.

The present bullseye grating was designed using a conventional forward parameter optimization. Since the grating primarily serves to convert otherwise non-radiative high-*k* modes into propagating free-space radiation, further improvements in outcoupling efficiency should be achievable through inverse-design or topology-optimization approaches. The enhancement reported here therefore represents a conservative estimate rather than the ultimate performance limit of the proposed architecture.

Unlike conventional directional emitters that concentrate radiation along the surface normal, the HMM redistributes the emitted power into the hyperbolic resonance cone supported by the multilayer. This produces the characteristic two-lobed angular emission profile shown in Figure 4b together with the annular intensity distribution observed at the detector plane (Figure 3c). Although the peak emission occurs away from the surface normal, the emitted power remains confined within a relatively narrow range of propagation angles, enabling efficient collection by conventional planar photodetectors while avoiding the nearly isotropic emission characteristic of bulk scintillators.

An important practical consideration is the robustness of the enhancement mechanism to fabrication-induced deviations in the Ag and $CsPbBr_3$ layer thicknesses. To evaluate this, we introduced random thickness variations of ±3 nm into every layer and calculated the scintillation enhancement for 51 randomly generated multilayer configurations using the single-dipole configuration described in Section S2 of the Supporting Information. Although modest variations in enhancement are observed, the average response remains close to the nominal design, demonstrating that the proposed architecture is tolerant to realistic fabrication deviations (see Supporting Information, Section S4).

Another practical consideration is whether the multilayer contains sufficient scintillator volume for efficient X-ray detection. The proof-of-concept device investigated here employs only seven Ag/$CsPbBr_3$ bilayers to maintain computationally tractable simulation times for the full dipole-ensemble calculations. In practice, however, the multilayer architecture can be readily extended to substantially larger numbers of periods. To verify this scalability, we simulated a 100-layer HMM containing one dipole located at the centre of every perovskite layer. As shown in Section S5 of the Supporting Information, the enlarged structure preserves both the characteristic ring-shaped detector-plane intensity distribution and the hyperbolic angular emission profile, confirming that the underlying enhancement mechanism remains effective even in substantially thicker multilayer architectures. Increasing the number of active scintillating layers therefore provides a practical route toward higher overall photon yields while retaining the desirable emission characteristics of the HMM.

## 3. Conclusion

In summary, we have demonstrated that embedding a scintillation medium within a hyperbolic metamaterial provides an effective strategy for simultaneously engineering spontaneous-emission dynamics and light extraction. The proposed Ag/$CsPbBr_3$ multilayer architecture, combined with a bullseye outcoupling grating, achieves broadband enhancement of the detected scintillation signal by up to fivefold relative to an equivalent bulk scintillator through the combined action of Purcell-enhanced spontaneous emission and directional energy transport along the hyperbolic resonance cone. Since the design is compatible with established thin-film deposition and nanofabrication techniques and is tolerant to realistic fabrication deviations, it offers a practical route toward experimental realization. More broadly, this work establishes hyperbolic metamaterials as a versatile platform for next-generation scintillators in medical imaging, radiation detection, high-energy physics, and related photon-starved sensing applications.

## 4. Methods:

All electromagnetic simulations were performed using the finite-difference time-domain (FDTD) method implemented in a commercial solver (Ansys Lumerical FDTD Solutions). The HMM consisted of seven Ag/$CsPbBr_3$ bilayers (14 alternating layers) with Ag and $CsPbBr_3$ layer thicknesses of 16 nm and 20 nm, respectively, over a lateral area of $2 \times 2\ \mu m^2$. This lateral size was chosen as a compromise between computational efficiency and accurate representation of the electromagnetic response of the multilayer. The effective permittivities of the HMM were calculated independently using effective-medium theory. For the field-distribution calculations presented in Figures 2c and 2d, a substantially larger number of multilayer periods was employed to eliminate finite-size reflections and more closely approximate an effectively infinite HMM.

The simulation domain was terminated on all sides using perfectly matched layer (PML) absorbing boundary conditions with the stabilized stretched-coordinate PML formulation. Spatial discretization employed the automatic non-uniform mesh with mesh accuracy level 3 (approximately 14 points per wavelength), together with an override mesh covering the entire HMM region. The maximum mesh sizes were 10 nm along the lateral directions (x and y) and 5 nm along the stacking direction (z). Conformal Variant 1 meshing was employed to improve the representation of the metal–dielectric interfaces.

The optical response of the HMM was evaluated using electric dipole sources embedded within the $CsPbBr_3$ layers. Unless otherwise specified, the dipoles were assumed to be vertically oriented to efficiently excite the TM-polarized high-*k* electromagnetic states supported by the HMM. For the scintillation simulations, each perovskite layer contained eleven dipoles: one fixed at the centre of the layer and ten additional dipoles randomly distributed throughout the layer. The dipoles were assigned random initial phases to represent the incoherent nature of scintillation emission following X-ray excitation. Each dipole source was assigned a base amplitude of 1 fW and a (dimensionless) amplitude of 1, such that the total source power was 1 fW for every dipole. Seventeen independent random realizations of the dipole ensemble were simulated, and all reported optical power densities and optical powers correspond to ensemble-averaged values. A convergence study (Supporting Information, Figure S8) showed that the ensemble-averaged enhancement factor had effectively converged after 17 independent realizations. Consequently, additional realizations were not performed, as they would incur a substantial computational cost while yielding negligible changes in the reported results.

The electromagnetic field distribution and the magnitude of the time-averaged Poynting vector were recorded using field and power monitors. A z-normal surface monitor spanning the entire *xy* plane and positioned 1 μm above the bullseye grating was used to calculate both the optical power density and the surface-integrated optical power reaching the detector. Unless otherwise stated, all reported results correspond to steady-state solutions at the wavelength of interest.

The germanium bullseye grating dimensions were optimized using a forward parameter sweep to maximize the optical power collected by the detector at the peak $CsPbBr_3$ emission wavelength of 523 nm. Optical constants of germanium were taken from literature.[39]

**Acknowledgements:**

The authors acknowledge support from the Department of Atomic Energy, Government of India, under Project Identification No. RTI 40007.

# Hyperbolic metamaterials for enhanced scintillation

**Priyankar Pandey[1], Subrahmanyam Mantha[1,2], Harish N S Krishnamoorthy[1*]**
[1] *Tata Institute of Fundamental Research Hyderabad*
[2] *Birla Institute of Technology and Science Pilani, KK Birla Goa Campus*
**Correspondence: harishk@tifrh.res.in*

## Supporting Information

### S1. Dielectric permittivities of individual layers

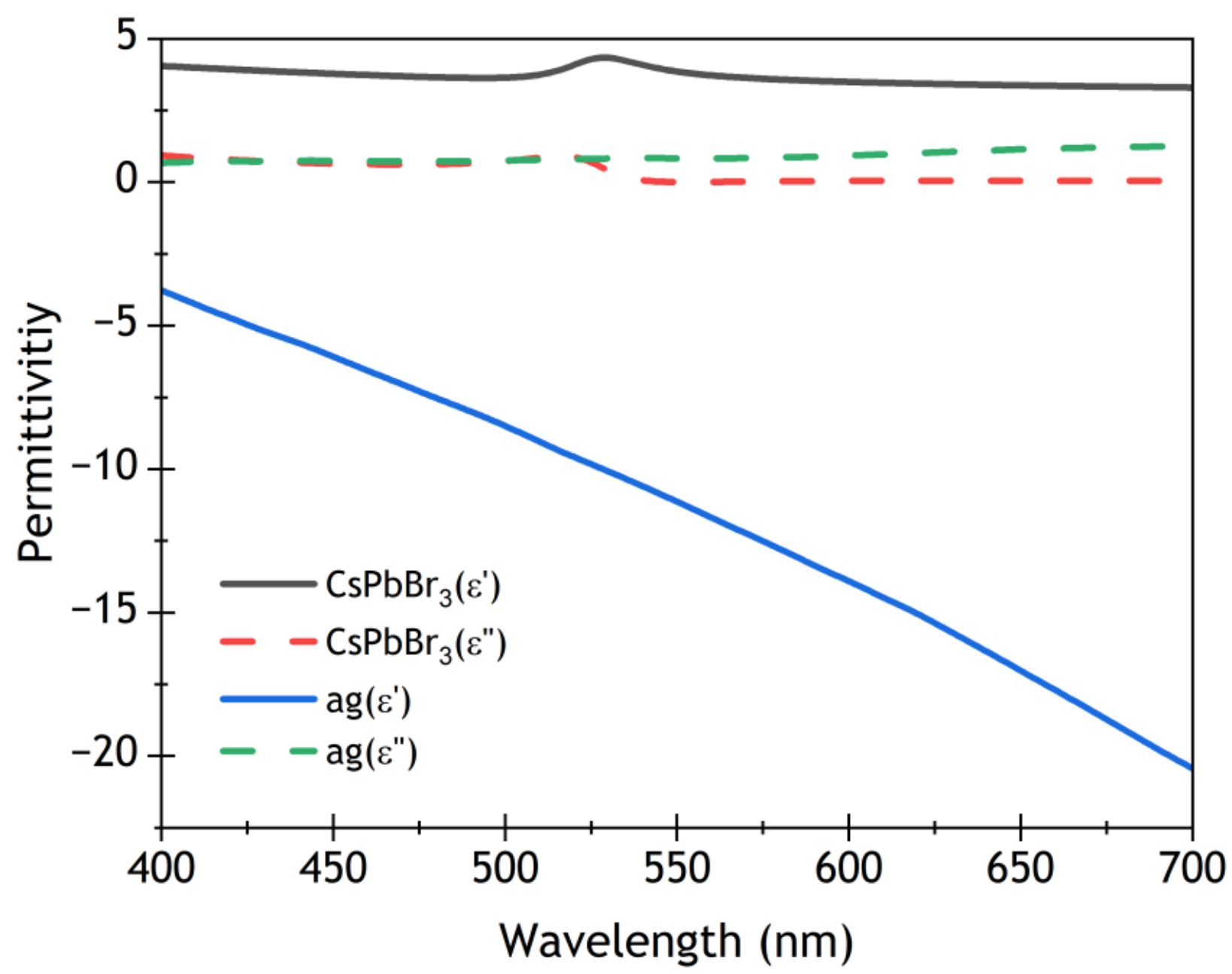

***Figure S1.*** *Real and imaginary parts of the dielectric permittivities as a function of wavelength of individual layers Ag and $CsPbBr_3$ that comprise the HMM taken from literature.*

### S2. Single dipole simulations

Single-dipole simulations were performed to elucidate the fundamental electromagnetic response of the HMM, including the Purcell enhancement of spontaneous emission and the directional emission characteristics arising from the hyperbolic dispersion. A single vertically oriented electric dipole was embedded at the centre of the fifth $CsPbBr_3$ layer of the HMM, corresponding to the dipole configuration used in the main text.

### S2.1. Purcell factor of the HMM structure

Figure S2 shows the Purcell factor, defined as the spontaneous-emission rate enhancement, calculated for a single vertically oriented electric dipole embedded at the centre of the fifth $CsPbBr_3$ layer of the HMM. The Purcell factor was obtained by calculating the total power emitted by the dipole in the HMM using a closed box of power monitors surrounding the source and normalizing it to the power emitted by an identical dipole embedded in a homogeneous $CsPbBr_3$ slab having the same overall thickness as the HMM. Since the spontaneous-emission rate of an electric dipole is proportional to its radiated power, the Purcell factor is given by:

$$F_P = \frac{P_{HMM}}{P_{Bulk}}$$

where $P_{\mathrm{HMM}}$ is the total power emitted by the dipole in the HMM and $P_{\mathrm{Bulk}}$ is the power emitted by the same dipole embedded in the homogeneous $CsPbBr_3$ reference structure.

The calculated Purcell factor exceeds 80 near the peak emission wavelength of $CsPbBr_3$, confirming the substantial enhancement of the local density of optical states supported by the HMM. The reduction in the Purcell factor at shorter wavelengths arises primarily from increased optical absorption in $CsPbBr_3$, where the imaginary part of the dielectric function is significantly larger.

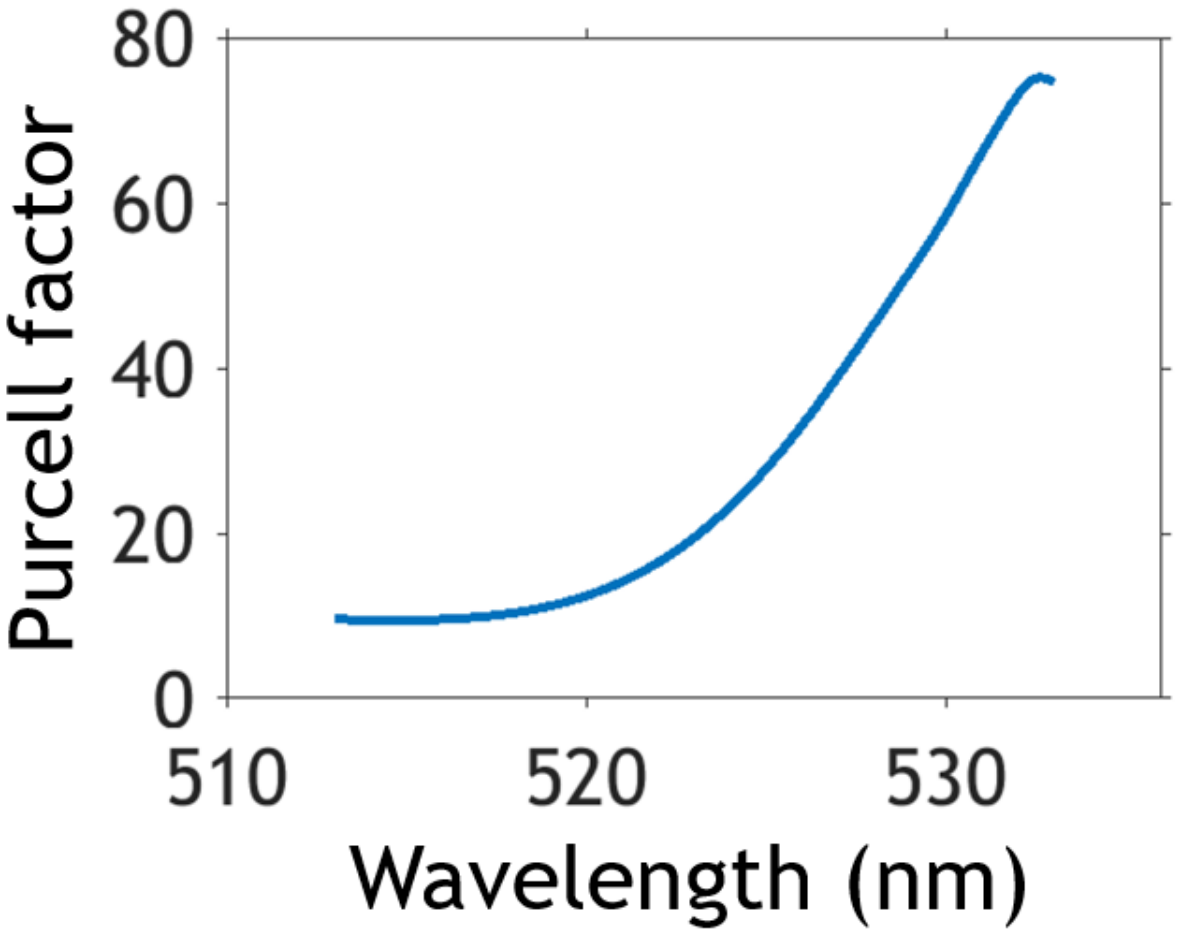

**Figure S2.** *Purcell factor (spontaneous-emission rate enhancement) as a function of wavelength for a single vertically oriented electric dipole embedded at the centre of the fifth $CsPbBr_3$ layer of the HMM.*

**S2.2. Emission profile of HMM scintillator**

To illustrate the effect of the HMM on the far-field emission characteristics, a single vertically oriented dipole emitting at the $CsPbBr_3$ photoluminescence peak ($\lambda = 523$ nm) was placed at the centre of the fifth perovskite layer. The resulting electromagnetic fields and optical power density were compared with those of a homogeneous $CsPbBr_3$ slab of identical thickness.

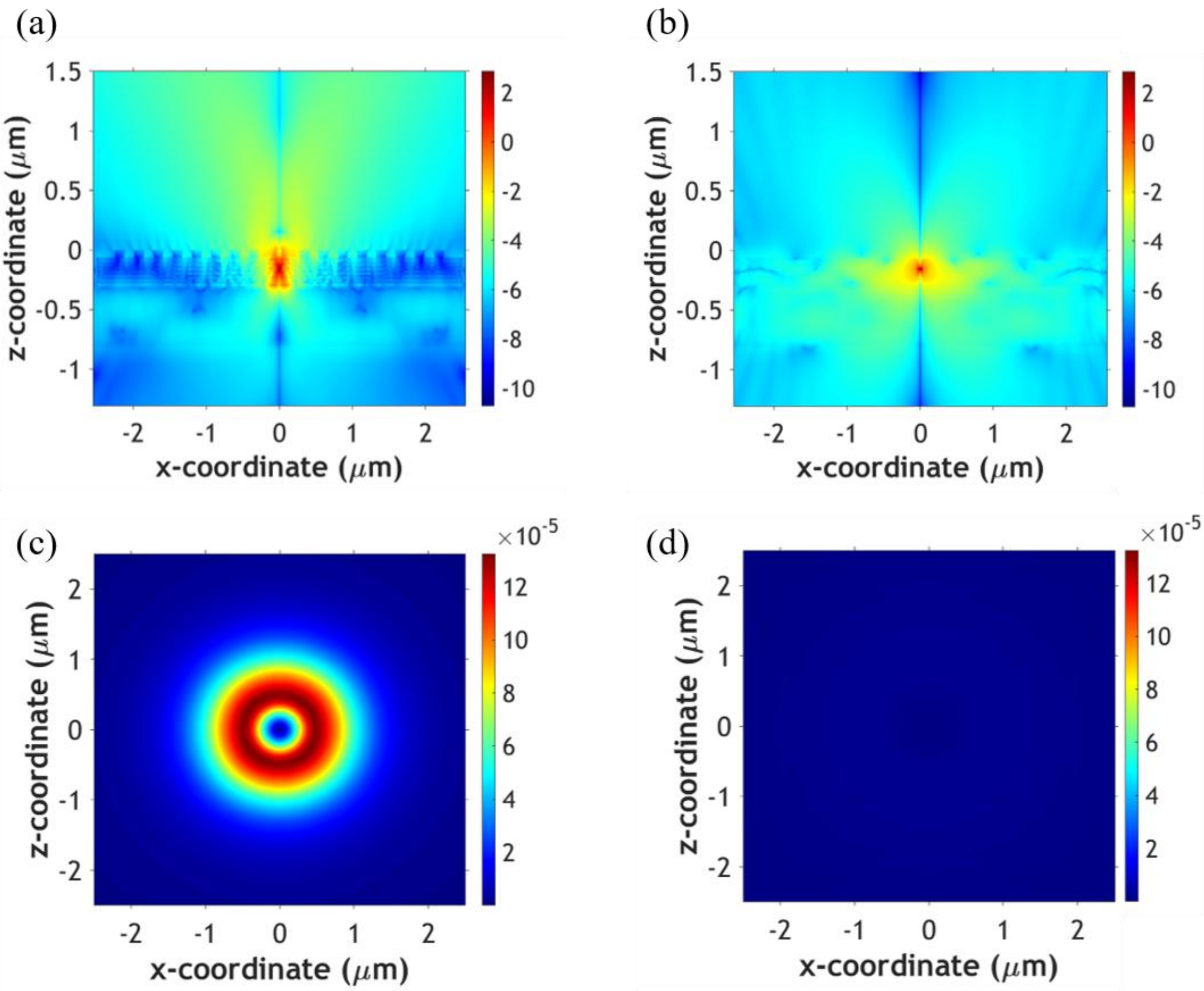

***Figure S3.*** *Colour maps of the magnitude of the time-averaged Poynting vector at $\lambda = 523nm$ for a single vertically oriented electric dipole. (a,b) Spatial distribution in the $xz$plane for the (a) HMM scintillator and (b) homogeneous bulk $CsPbBr_3$ reference. The HMM exhibits highly directional emission associated with the hyperbolic resonance cone, whereas the bulk scintillator produces weak, nearly isotropic radiation. (c,d) Optical power density (magnitude of the time-averaged Poynting vector) recorded at a detector plane located 1 μm above the structures for the (c) HMM and (d) bulk configurations. The HMM produces a pronounced annular (ring-shaped) intensity distribution, characteristic of conical emission from the hyperbolic resonance cone, and delivers substantially greater optical power to the detector than the bulk scintillator. Compared with the ensemble-dipole simulations presented in Figure 3 of the main text, the ring pattern is considerably sharper because it originates from the emission of a single dipole rather than the incoherent superposition of multiple emitters with random positions and phases.*

### S2.3. Scintillation enhancement for single dipole

Figure S4 shows the scintillation enhancement factor, $\eta$, as a function of wavelength for the single-dipole configuration. In contrast to the dipole-ensemble results presented in the main text (Figure 4a), the enhancement spectrum exhibits a more symmetric variation about the $CsPbBr_3$ emission peak. This behaviour confirms that the apparent asymmetry observed for the dipole ensemble arises primarily from the wavelength dependence of the bulk reference signal, which becomes very small at shorter wavelengths and consequently inflates the ratio $\eta$.

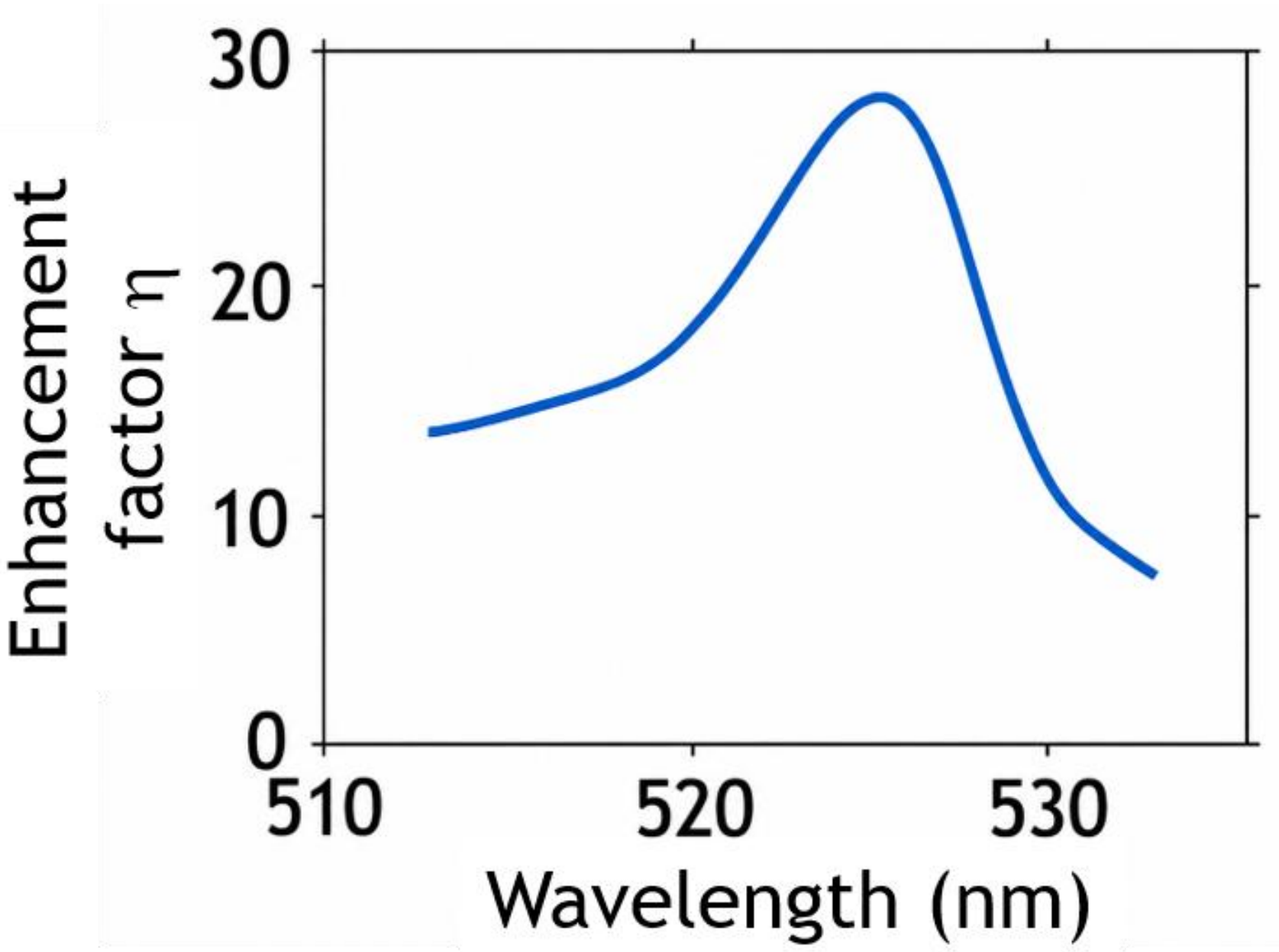

***Figure S4.*** *Spectral variation of the scintillation enhancement factor,* $\eta = P_{HMM}/P_{Bulk}$*, evaluated at the detector plane located 1* $\mu m$ *above the structure for the single-dipole configuration. The enhancement reaches approximately fivefold at the* $CsPbBr_3$ *emission wavelength (523 nm). Unlike the dipole-ensemble results shown in Figure 4a of the main text, the enhancement spectrum is nearly symmetric about the peak, confirming that the asymmetry observed for the ensemble case originates primarily from the wavelength dependence of the bulk reference signal rather than from the intrinsic response of the HMM.*

## S3. Power spectra for dipole ensemble

Figure S5 compares the optical power density reaching the detector for the HMM and the homogeneous bulk $CsPbBr_3$ reference under the dipole-ensemble configuration. The HMM delivers substantially higher optical power density across the entire emission band. At shorter wavelengths, the optical power density of the bulk reference becomes particularly small, causing the scintillation enhancement factor $\eta$ to become more sensitive to the denominator. Consequently, the enhancement spectrum shown in Figure 4a of the main text exhibits a slight asymmetry on the shorter-wavelength side, which is a normalization effect rather than an intrinsic feature of the HMM response.

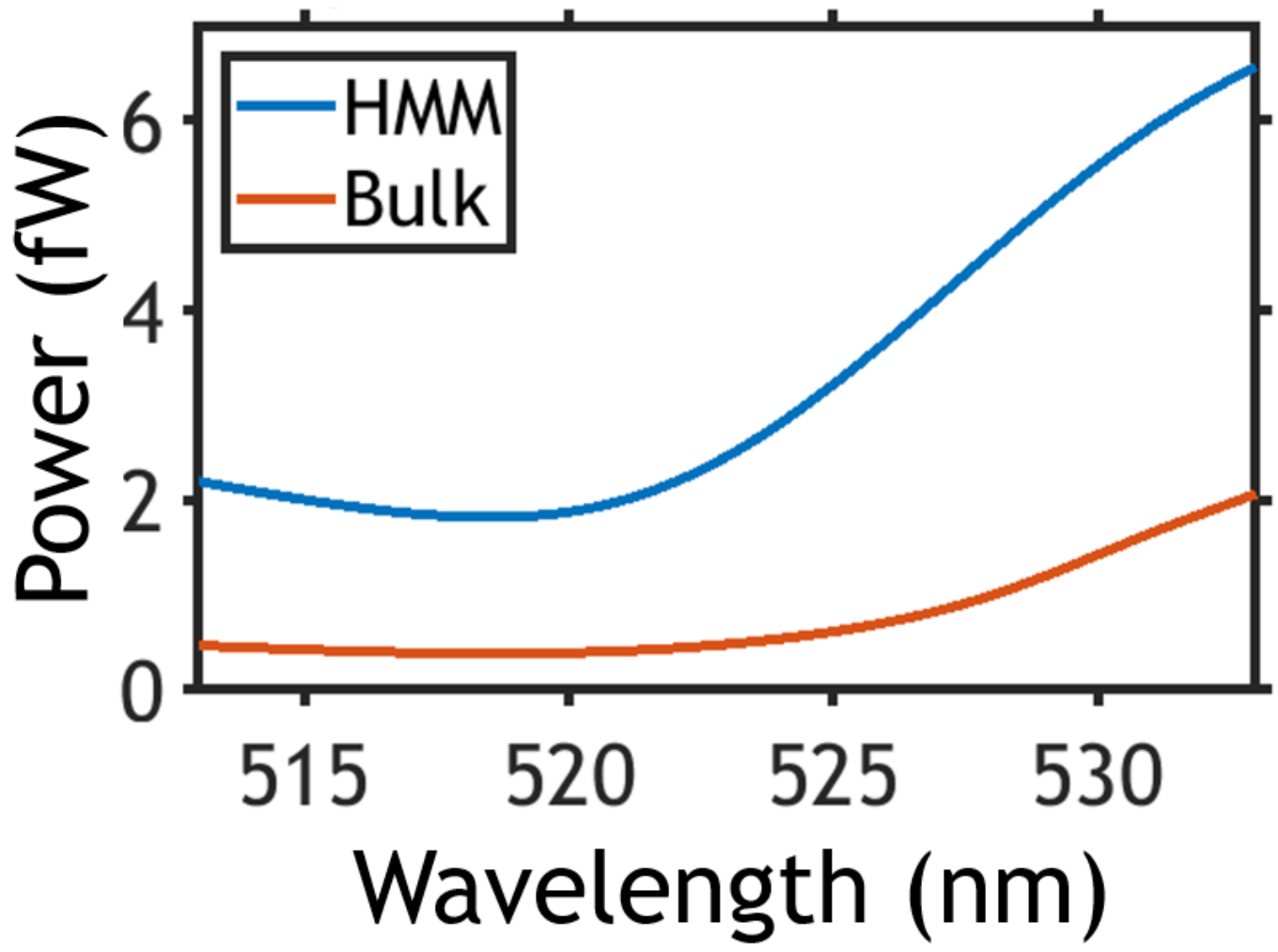

***Figure S5.*** *Spectral variation of the optical power density (magnitude of the time-averaged Poynting vector) evaluated at the detector plane located 1 μm above the structure for the HMM (blue) and the homogeneous bulk* $CsPbBr_3$ *reference (orange) under the dipole-ensemble configuration. The HMM exhibits substantially higher optical power density throughout the emission band, while the comparatively small optical power density of the bulk reference at shorter wavelengths contributes to the apparent asymmetry in the scintillation enhancement spectrum.*

**S4. Fabrication tolerance analysis of the enhancement factor**

To evaluate the robustness of the proposed HMM scintillator against fabrication imperfections, a fabrication-tolerance analysis was performed using the single-dipole configuration. A vertically oriented electric dipole was positioned at the centre of the fifth $CsPbBr_3$ layer, identical to the configuration employed for the Purcell-factor calculations in Section S2.1. The single-dipole model was adopted because it captures the underlying electromagnetic response of the HMM while enabling the evaluation of a sufficiently large number of perturbed multilayer geometries within practical computational limits. Random thickness variations of ±3 nm were introduced into every Ag and $CsPbBr_3$ layer using the built-in pseudorandom number generator in Ansys Lumerical, while preserving the nominal multilayer architecture. A total of 51 randomly generated multilayer configurations were simulated, and the scintillation enhancement factor was calculated for each configuration.

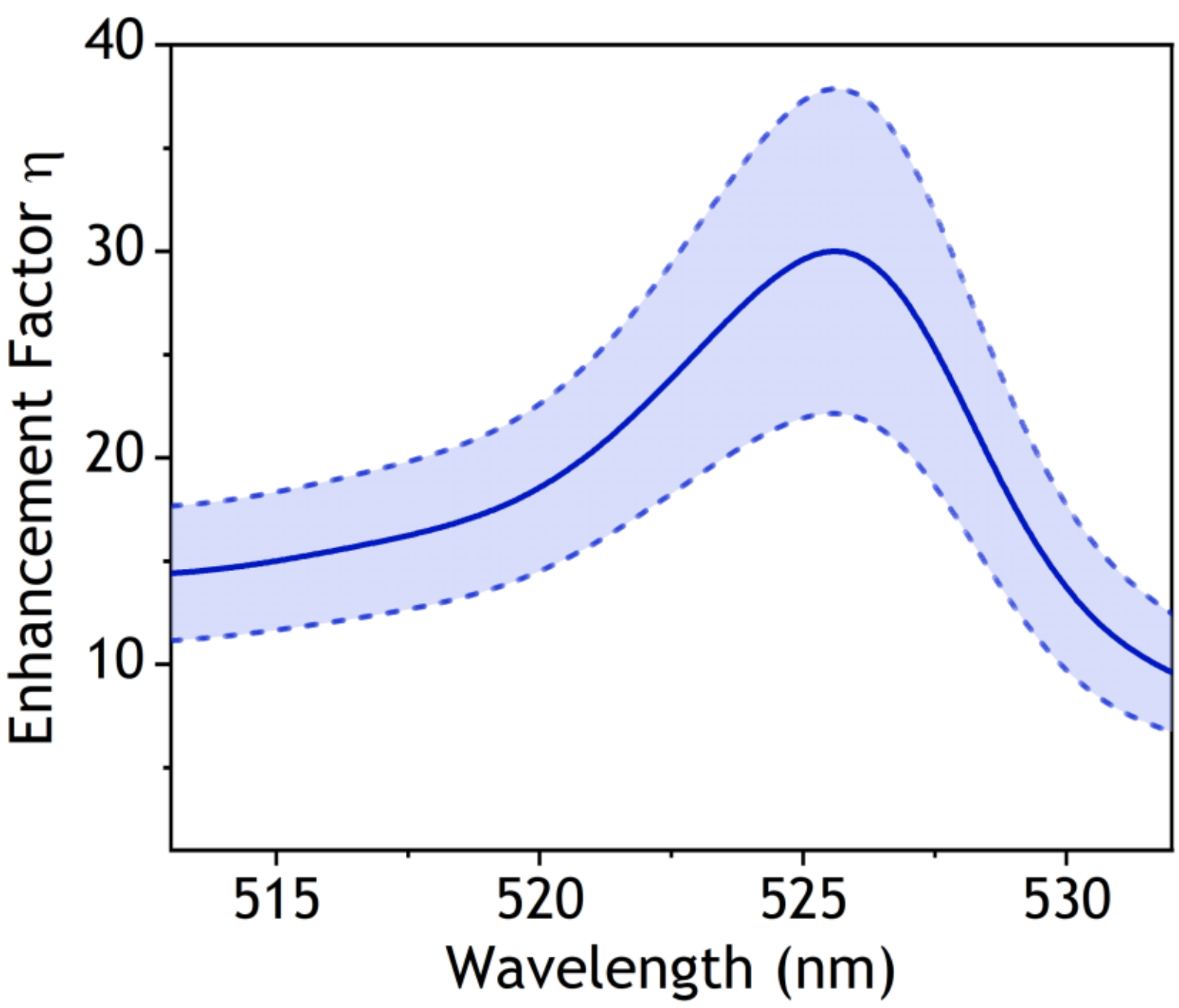

***Figure S6.*** *Mean enhancement factor (solid line) with the corresponding ±1 standard deviation (shaded region) obtained from 51 randomly generated multilayer realizations incorporating random thickness variations of up to ±3 nm in every Ag and $CsPbBr_3$ layer. The calculations were performed for a single vertically oriented dipole located at the centre of the fifth $CsPbBr_3$ layer. The small variation about the mean demonstrates that the enhancement mechanism is robust against realistic fabrication-induced thickness errors.*

**S5. Scalability of the HMM scintillator**

To investigate whether the proposed enhancement mechanism persists in thicker structures that are more representative of practical scintillators, additional simulations were performed for an HMM comprising 100 alternating Ag and $CsPbBr_3$ layers. To maintain computational feasibility, one vertically oriented electric dipole was positioned at the centre of each $CsPbBr_3$ layer. This excitation model enables assessment of the optical response of the larger multilayer while avoiding the substantial computational cost associated with ensemble averaging over many randomly distributed dipoles.

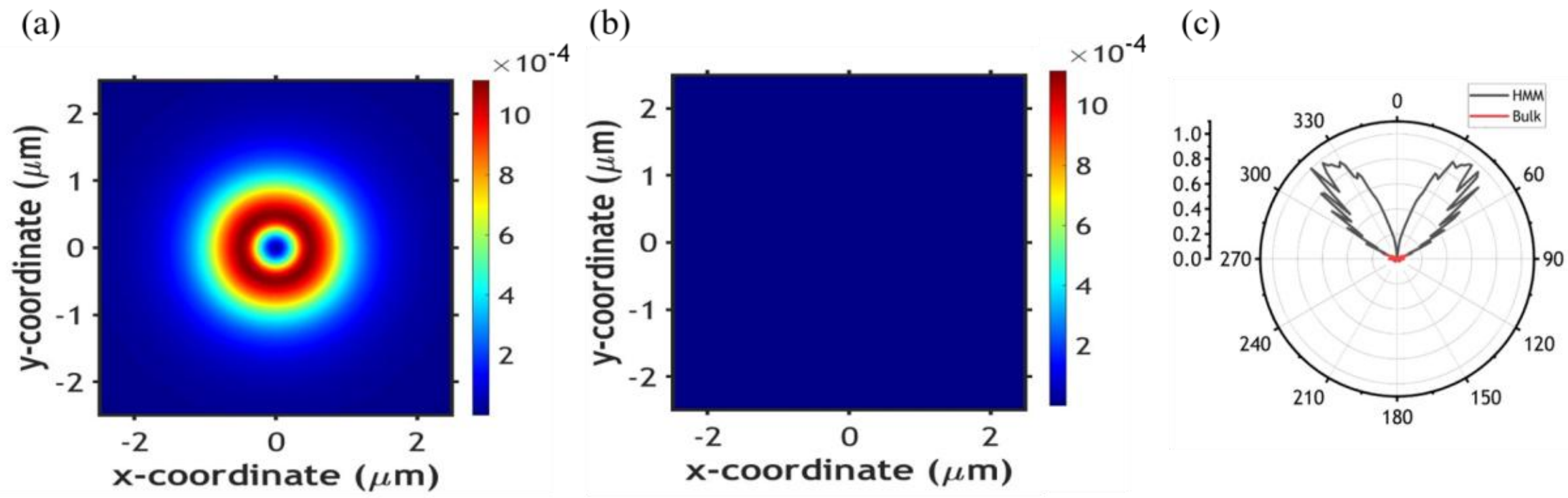

***Figure S7.*** *Scalability of the proposed HMM scintillator to larger multilayer structures. (a) Optical power density at the detector plane located 1 μm above a 100-layer Ag/CsPbBr₃ HMM excited by one vertically oriented electric dipole positioned at the centre of each perovskite layer. The characteristic annular emission pattern is preserved despite the substantially increased number of layers. (b) Corresponding optical power density for a homogeneous bulk CsPbBr₃ scintillator of identical thickness under identical excitation conditions. Both panels are plotted using the same colour scale. (c) Far-field angular emission profiles for the HMM and bulk scintillator. The HMM retains the characteristic two-lobed angular distribution associated with the hyperbolic resonance cone, demonstrating that the underlying enhancement mechanism is preserved in substantially thicker multilayer structures.*

**S6. Convergence of dipole-ensemble averaging**

To establish the statistical convergence of the dipole-ensemble model, the enhancement factor was evaluated by progressively increasing the number of independent random dipole realizations. Each realization consisted of randomly distributed, vertically oriented dipoles within the $CsPbBr_3$ layers, with one dipole fixed at the centre of every perovskite layer to ensure uniform excitation of all scintillating layers. The enhancement spectrum was calculated after ensemble averaging over 5, 8, 11, 14, and 17 independent realizations.

As shown in Figure S8, the enhancement spectrum converges rapidly with increasing ensemble size. While noticeable variations are present for a small number of realizations, the spectral response changes only marginally beyond approximately 14–17 realizations, indicating that 17 independent realizations are sufficient to obtain statistically converged enhancement spectra for the parameters considered in this work.

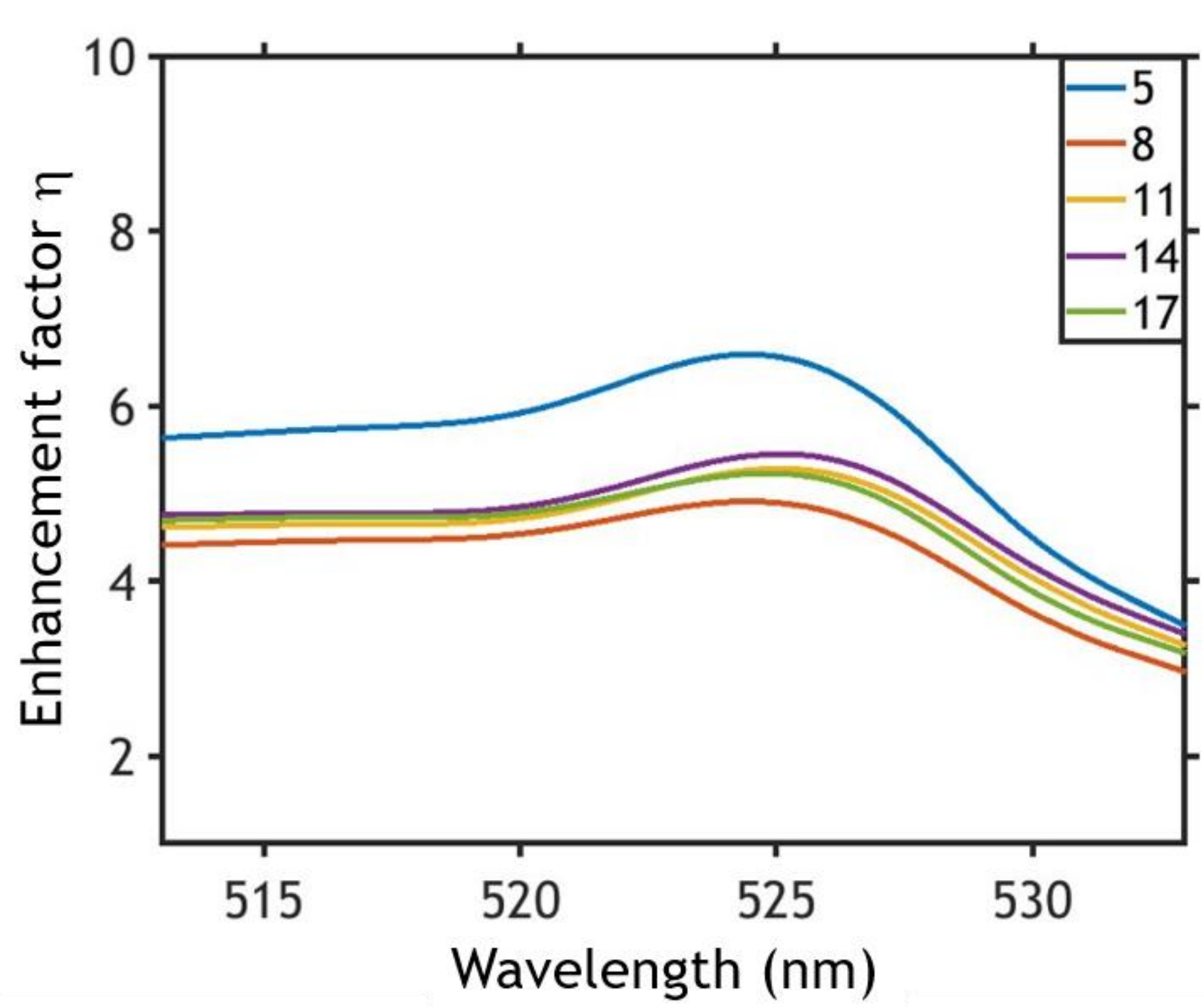

***Figure S8.*** *Convergence of the calculated enhancement factor with increasing number of independent dipole-ensemble realizations. The enhancement spectrum is shown after ensemble averaging over 5, 8, 11, 14, and 17 independent realizations. The small change in the spectrum beyond 14–17 realizations indicates that the calculated enhancement has reached statistical convergence.*